\documentclass[prb,aps,twocolumn,floats,showpacs]{revtex4}
\usepackage{graphicx}
\usepackage{amsmath}

\renewcommand{\eqref}[1]{Eq.~(\ref{#1})}

\newcommand{\non}{\nonumber}
\newcommand{\be}{\begin{equation}}
\newcommand{\ee}{\end{equation}}
\newcommand{\bea}{\begin{eqnarray}}
\newcommand{\eea}{\end{eqnarray}}
\newcommand{\bse}{\begin{subequations}}
\newcommand{\ese}{\end{subequations}}

\newcommand{\lb}{\left[}
\newcommand{\rb}{\right]}
\newcommand{\lp}{\left(}
\newcommand{\rp}{\right)}

\newcommand{\p}{\mathbf{p}}

\newcommand{\E}{{\cal E}}

\newcommand{\sign}{\mbox{\,sign\,}}

\newcommand{\half}{\textstyle{\frac12}}
\renewcommand{\j}{j}

\begin{document}

\title{Numbers of donors and acceptors from transport measurements in graphene}
\author{D.~S.~Novikov}
\affiliation{W. I. Fine Institute of Theoretical Physics, 
University of Minnesota, Minneapolis, MN 55455, USA}
\date{\today}


\begin{abstract}

A method is suggested to separately determine the surface density of positively
and negatively charged impurities that limit the mobility in a graphene
monolayer. The method is based on the exact result for the transport cross-section, 
according to which the massless carriers  
are scattered more strongly when they are 
attracted to a charged impurity than when they are repelled from it.

\end{abstract}

\maketitle

The discovery of graphene\cite{discovery} 
opened up a number of research directions.\cite{rise}
The outstanding properties of this material originate from 
its massless electron-hole symmetric carrier dispersion
$\epsilon(\p)=\pm vp$, where the Fermi velocity $v\approx 10^6\,$m/s.
\cite{discovery,rise,novoselov,deheer,zhang} 
Recent transport measurements \cite{novoselov,deheer,zhang} show 
that the graphene mobility is approximately independent of the carrier density. 
It is believed that the mobility is limited by  
the Coulomb impurities in the substrate, as well as by the
quenched ripples.\cite{corrugations} 
In general, the carrier density and the mobility are sensitive to 
the immediate environment, e.g. to the presence of the adsorbed molecules, 
or to the donors and acceptors in the substrate. This has prompted proposals 
for applications of graphene monolayers as gas sensors.\cite{sensors}
Models\cite{ando'06,nomura,hwang-adam-dassarma,ostrovsky} based on the Born 
approximation for scattering off the Coulomb impurities,
suggest a natural way to estimate the surface impurity density $n_i$.

In a typical situation, the Coulomb impurity density $n_i = n_i^+ + n_i^-$
consists of both positively and negatively charged species,
here referred to as donors and acceptors. 
In this work I suggest a means to determine the densities $n_i^\pm$ 
{\it separately} from a dc conductivity as a function of the 
gate voltage. This may find applications in the sensing of adsorbed gas molecules,
that can be either donors or acceptors,\cite{sensors} 
as well as in the characterization of the substrate surface.

The proposed effect is based on the recent observation\cite{novikov}
that the exact transport cross-section is fairly sensitive to whether the 
charge carrier is attracted to an impurity or is repelled from it. 
For instance, for a conduction electron the transport cross-section is greater 
when scattering off a donor than off an acceptor.
This attraction-repulsion asymmetry is an unexpected feature for scattering
in the  $1/r$-potential, and is specific to the ``relativistic'' 
quasiparticle dispersion.
Indeed, such an asymmetry is absent in the exact nonrelativistic scattering
off a $1/r$ potential in two\cite{Stern-Howard} and three\cite{Landau3} dimensions,
which is why the proposed effect would not work for conventional semiconductors.
Remarkably, the asymmetry is also absent for scattering both off local (neutral)
imperfections\cite{katsnelson-novoselov} and off
quenched corrugations.\cite{corrugations}
Thus the proposed method selectively senses the imbalance 
between the {\it charged} impurities.


Consider the scattering off the Coulomb potential
\be \label{U-coul}
U(r) = -{Ze_*^2\over r} 
\equiv - \hbar v \times {\alpha_0 \over r} 
\ee
where the dimensionless impurity strength 
\be \label{alpha}
\alpha_0 = {Z e_*^2/\hbar v} \,, \quad e_*^2 ={2e^2/(\varepsilon+1)} 
\ee
can be both positive (attraction) and negative (repulsion). 
Here $Z$ is the impurity valence, and $\varepsilon$ is the dielectric constant
of a substrate.
The vacuum value $\alpha_0|_{Z=1, \varepsilon=1}\approx 2.2$ 
for $v\approx 1\times 10^6\,$m/s, while for the SiO$_2$ substrate,
$\alpha_0|_{Z=1, \varepsilon=3.9}\approx 0.9$.
The effects of interactions between carriers in graphene,
estimated via the scale-invariant RPA screening,\cite{screening} 
further diminish $\alpha_0$, 
\be \label{RPA}
\alpha_0 \to \alpha = \alpha_0/\varepsilon_{\rm RPA}, 
\quad \varepsilon_{\rm RPA}  
= 1+ (\pi/2)\times {e_*^2/\hbar v} \,.
\ee
This reduces the impurity strength by the factor
$\varepsilon_{\rm RPA}|_{\varepsilon=3.9}\approx 2.4$
for a monovalent impurity on a SiO$_2$ substrate, yielding 
$|\alpha|\approx 0.37$.

\begin{figure}[t]
\begin{minipage}[t]{3.6in}
\includegraphics[width=3.6in]{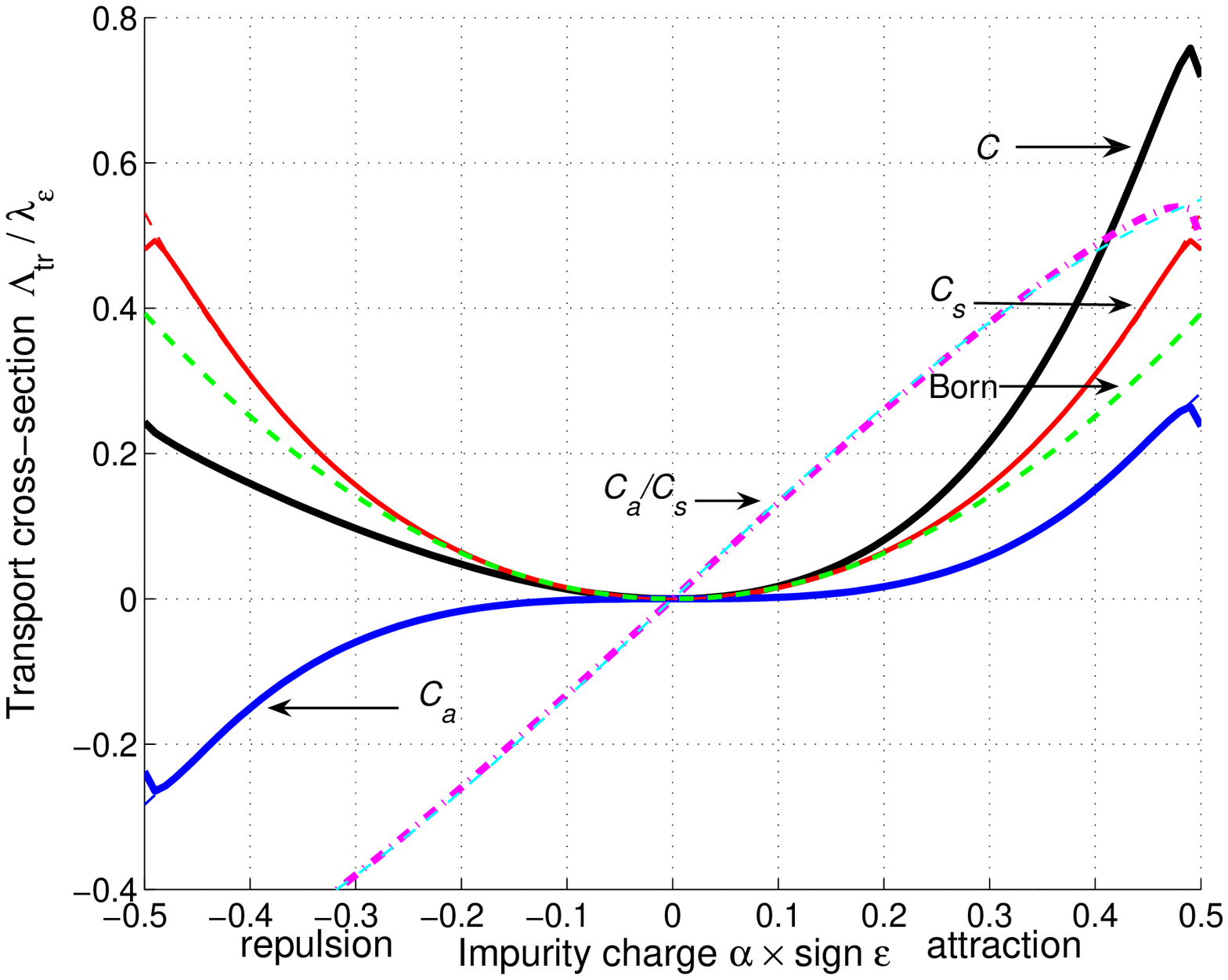}
\end{minipage}
\begin{minipage}[t]{1.65in}
\vspace{-3in}\hspace{-0.21in}
\includegraphics[width=1.65in]{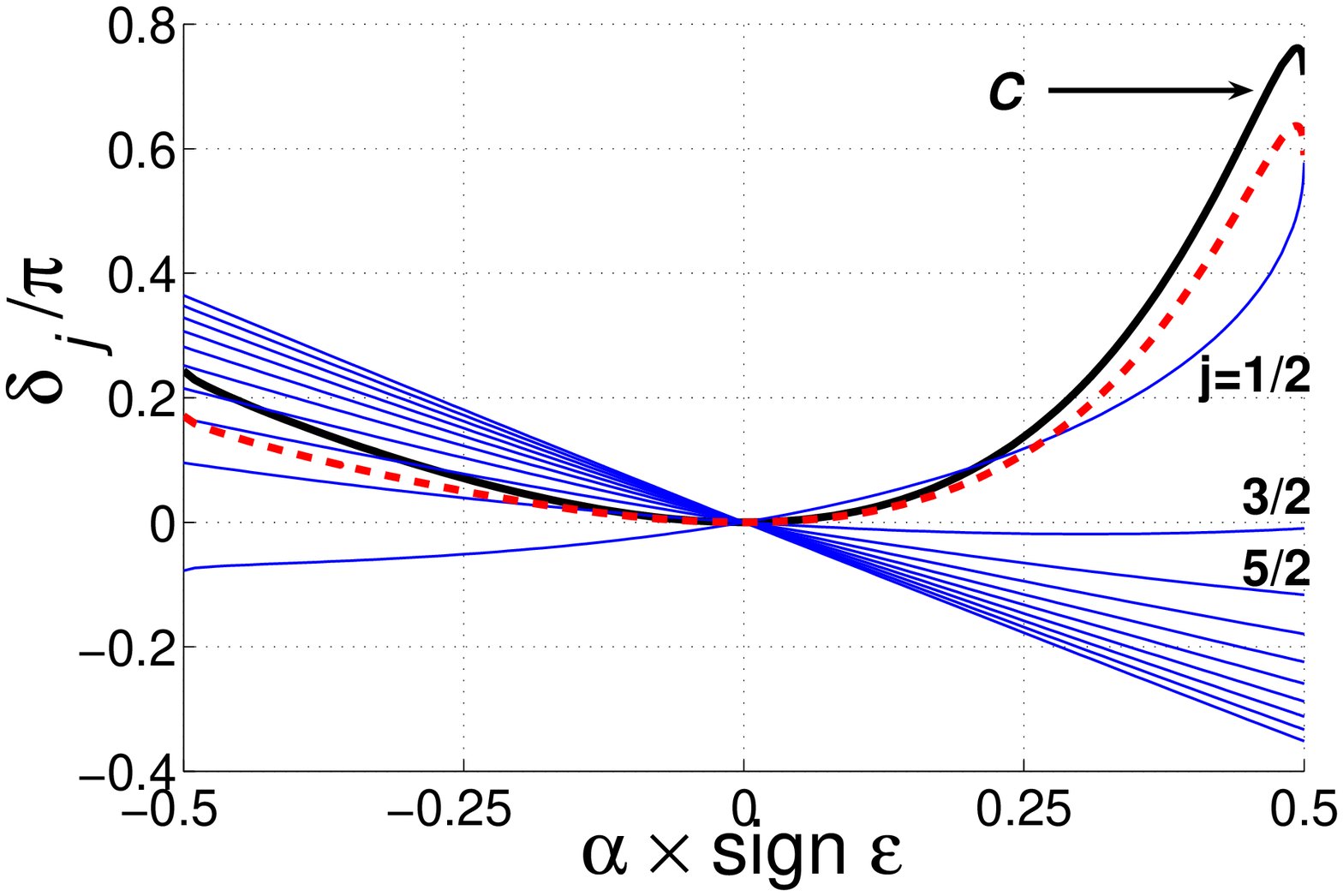}
\end{minipage}
\caption{(Color online) 
Even, $C_s(\alpha_\epsilon)$ (red), and odd, $C_a(\alpha_\epsilon)$ (blue), 
parts of the exact transport cross-section $C(\alpha_\epsilon)$ 
[Eq.~(\ref{lambda-tr-C})] (black), their ratio $C_a/C_s$ (dash-dot),
together with the polynomial fits (\ref{C-fit}) (dashed). 
Also shown is the Born approximation 
$C^{\rm Born} = {\pi\over 2}\alpha_\epsilon^2$ (green dashed).
{\it Inset:} First ten scattering phase shifts $\delta_j(\alpha_\epsilon)$,
$j=\frac12 ... \frac{19}2$. Note that $\delta_{1/2}$
behaves very differently from the rest. The differences 
$\delta_{j+1}-\delta_j$ decay as $1/j$.\cite{stirling}
Including only $\delta_{1/2}$ and $\delta_{3/2}$  
already gives a good approximation (red dashed) to the exact sum $C$ (black). 
  }
\label{fig:Csa}
\end{figure}

The scattering states for the problem (\ref{U-coul}) are characterized by 
the angular momentum $j=m+\half$, $m=0,\pm1,\pm2, ...$ .
The scattering phase shifts $\delta_j$ are\cite{novikov}
\be \label{S-coul-ultra}
e^{2i\delta_j} =  {\j \, e^{i\pi(\j-\gamma)} \over \gamma- i\alpha_\epsilon} 
{\Gamma(1+\gamma - i\alpha_\epsilon)\over 
\Gamma(1+\gamma +i\alpha_\epsilon)} \,, 
\quad  \alpha_\epsilon \equiv \alpha \sign \epsilon \,.
\ee
Here $\gamma=\sqrt{j^2-\alpha^2}$, 
and $|\alpha|$ in this formula is limited by $\half$
(the subcritical impurity\cite{novikov,supercrit}).
The property $\delta_j = \delta_{-j}$ (mod $\pi$) 
ensures the absence of backscattering.
Since the phase shifts depend only on $\sign \epsilon$ but not on the 
absolute value $|\epsilon|$ of the quasiparticle energy, 
the exact transport cross-section
\be \label{lambda-tr-C}
\Lambda_{\rm tr} = C(\alpha_\epsilon) \times \lambda_\epsilon \,, 
\quad C= {2\over\pi}
\sum_{j=1/2}^\infty
\sin^2\lp \delta_{j+1}-\delta_j \rp
\ee
is proportional to the energy-dependent carrier wavelength 
$\lambda_\epsilon={2\pi\hbar v/|\epsilon|}$.
The dimensionless function $C(\alpha_\epsilon)$,
which is the transport cross-section in the units of the carrier wavelength, 
is plotted in Fig.~\ref{fig:Csa}.

The transport cross-section (\ref{lambda-tr-C})
is strongly asymmetric with respect to 
$\sign \alpha \times \sign \epsilon$, i.e. the sign of the potential
as seen by the carrier: A  donor ($\alpha>0$)
scatters conduction electrons ($\epsilon>0$) more effectively 
than it scatters holes $(\epsilon<0)$.
The attraction-repulsion asymmetry is commonplace 
[e.g. Problem 6 to Sec.~132 in Ref.~\onlinecite{Landau3}].
Physically, one may expect the particle to spend more time around an attractive
potential center and thereby be more significantly deflected
(this intuition, strictly speaking, applies to the massive particles).
However, for the practically important 2D and 3D Coulomb scattering in a parabolic
band, the corresponding exact solutions are somewhat exceptional in a sense
that they lack such an asymmetry. Remarkably, for the ``relativistic'' carrier
dispersion, characteristic of graphene, this generally expected asymmetry
is recovered.

The result (\ref{lambda-tr-C}) applies 
to the half-filled $\pi$-electron band. In this case the RPA screening 
(\ref{RPA}) is scale-invariant, preserving the {functional form} 
of the potential (\ref{U-coul}). 
At finite carrier density $k_F^2/\pi$, corresponding to the Fermi momentum $k_F$,
the RPA screening is not scale-invariant anymore.
\cite{ando'06,hwang-adam-dassarma} This means that the screened potential
has the form (\ref{U-coul}) only at distances shorter than the screening 
length $l_s \simeq Z/\alpha k_F$, and is cut-off at larger distances.
Scattering in such an effective potential is not tractable exactly,
and the momentum relaxation rate is practically calculated only
in the Born approximation.
\cite{ando'06,nomura,hwang-adam-dassarma,ostrovsky}
The latter becomes asymptotically exact 
for small $|\alpha|\ll 1$ at large distances $r\sim l_s\gg 1/k_F$,
but misses non-perturbative effects at distances shorter than $l_s\sim 1/k_F$
when the potential is large, $\alpha\sim 1$.

To investigate the role of the asymmetry, below I 
neglect the additional screening at finite $k_F$,
and use the scale-invariant screening (\ref{RPA}) by 
the filled valence band only. Such an approach may be justified noting 
that, while the divergent Coulomb phase\cite{Landau3} $\ln 2k_F r$ 
(same for all the angular momentum channels) 
comes from the distances $k_F r\gg 1$, the relative scattering phases $\delta_j$, 
that contribute to the momentum relaxation via the transport cross-section,
accumulate on the scale  $l_j = j/k_F$.\cite{approach} 
This way the lowest phase shifts $\delta_j$ are practically acquired within 
the screening length $l_s\gtrsim l_j$, 
while the rest are strongly reduced by the screening
at finite $k_F$. Fortunately, the dominant contribution to the 
cross-section (\ref{lambda-tr-C}) comes\cite{stirling} from 
the channels with $j\sim 1$ (Fig.~\ref{fig:Csa} inset), 
that are least damaged by the additional screening at finite $k_F$. 
In particular, the lowest shift $\delta_{1/2}$ 
is behaving qualitatively differently from the rest, causing the pronounced
attraction-repulsion asymmetry. Such an anomalous behavior is to be expected,
as this is the channel in which the criticality, $|\alpha|=j$, 
is first reached.

To calculate the conductivity, 
consider for simplicity the case when all the donors have the same 
valence and are characterized by the strength $+\alpha$,
while all the acceptors have the opposite valence and the strength $-\alpha$;
hereon $\alpha>0$. 
Let $f_{\pm}(\p)$ be the distribution functions
for the particles and holes correspondingly.
With the electron interaction effects approximated via the screening (\ref{RPA}), 
the kinetic equations for each carrier type decouple,
\be \label{KE}
e\vec{\E} \cdot {\partial f_+\over \partial \p} 
= - {\delta f_+\over \tau_+(\epsilon_\p)} \,, \quad
-e\vec{\E} \cdot {\partial f_-\over \partial \p} 
= - {\delta f_-\over \tau_-(\epsilon_\p)} \,, 
\ee
where $e<0$ is the electron charge, and 
\be \label{tau}
\tau_+^{-1} = v \lb n_i^+ \Lambda_{\rm tr}^+ + n_i^- \Lambda_{\rm tr}^-\rb,\quad
\tau_-^{-1} = v \lb n_i^+ \Lambda_{\rm tr}^- + n_i^- \Lambda_{\rm tr}^+\rb.
\ee
Here $\Lambda_{\rm tr}^\pm = \lambda_\epsilon\times C(\pm \alpha)$
are the right and left parts of the cross-section in Fig.~\ref{fig:Csa}.
The rates (\ref{tau}) have straightforward meaning: The conduction
electrons scatter off the donors with the enhanced cross-section 
$\Lambda_{\rm tr}^+$ and off the acceptors with the reduced cross-section
$\Lambda_{\rm tr}^-$, while for the holes the situation is reversed.
The Coulomb transport times (\ref{tau}) are proportional to the quasiparticle energy,
\be
\begin{split} \label{tau-lambda}
\tau_+(\epsilon) = {|\epsilon|\over 2\pi\hbar v^2} 
\times {1\over n_i^+ C(+\alpha) + n_i^- C(-\alpha)} \,, \\
\tau_-(\epsilon) = {|\epsilon|\over 2\pi\hbar v^2} 
\times {1\over n_i^+ C(-\alpha) + n_i^- C(+\alpha)} \,.
\end{split}
\ee
From the kinetic equations (\ref{KE}) find the deviations
\be \label{deltaf}
\delta f_\pm = \pm e(\vec{\E}\hat\p) v \tau_\pm(\epsilon) \times
\lb - {\partial_\epsilon f_\pm^{(0)} }\rb ,
\quad \hat \p \equiv {\p/p}
\ee
of the distribution functions from the equilibrium Fermi distribution
$f_\pm^{(0)}(\epsilon) = 1/[e^{(\epsilon \mp \mu)/T}+1]$, where 
$\mu$ is chemical potential relative to the half-filled $\pi$ band.
The resulting electric current 
[here $N_f=2_{\rm spin}\times 2_{\rm valley}=4$ 
independent polarizations]
\[
e{\bf J} = e N_f v \int\! {d^2\p\over (2\pi\hbar)^2} \,
\hat\p\lp\delta f_+ -\delta f_-\rp \equiv \sigma \vec{\E} 
\]
corresponds to the dc conductivity
\be \label{cond-gen}
\sigma = {N_f e^2\over h}  
\int_0^\infty \! \epsilon d\epsilon 
\lb {\tau_+(\epsilon)\over 2\hbar}
\lp -{\partial_\epsilon f_+^{(0)} } \rp 
+ {\tau_-(\epsilon)\over 2\hbar}
\lp -{\partial_\epsilon f_-^{(0)} } \rp \rb .
\ee
Last, I express the conductivity (\ref{cond-gen}) via the carrier number
densities. 
Using integration by parts, the electron and hole densities $n^\pm$
can be cast in a similar form,
\be \label{n}
n^\pm(\mu) =  \int\! {N_f\, d^2\p\over (2\pi\hbar)^2}\, f_\pm^{(0)}(\epsilon_\p) 
= {N_f\over 4\pi\hbar^2 v^2} 
\int_0^\infty \! \epsilon^2 d\epsilon \lb -{\partial_\epsilon f_\pm^{(0)} } \rb.
\ee
Combining Eqs.~(\ref{tau-lambda}), (\ref{cond-gen}), and (\ref{n}), 
the conductivity 
\be \label{sigma-C}
\sigma = 
 {(e^2/h) n^+ \over n_i^+ C(\alpha) + n_i^- C(-\alpha)} 
+{(e^2/h) n^- \over n_i^+ C(-\alpha) + n_i^- C(\alpha)}  \,.
\ee

The asymmetry of the cross-section $C(\alpha)$ translates into the asymmetry
in the dependence of the conductivity (\ref{sigma-C}) on the 
net carrier density $n=n^+ + n^-$. Introducing the 
charge carrier imbalance $\delta n = n^+ - n^-$, 
as well as the symmetric and antisymmetric parts of the cross-section
\be \label{Csa}
C_{s,a}(\alpha) = \half \lb C(\alpha) \pm C(-\alpha)\rb ,
\ee
one can represent the conductivity (\ref{sigma-C}) in the form
\be \label{sigma-Csa}
\sigma = {e^2\over h} {n\over n_i C_s} \times
{1- (\delta n/n) c \over 1-c^2} \,, 
\quad c(\alpha) = {C_a\over C_s}\times {n_i^+-n_i^-\over n_i} \,.
\ee
Here $n_i=n_i^+ + n_i^-$ is the total surface density of charged impurities.
Thus measuring the conductivity asymmetry for,
say, opposite gate voltages relative to the charge neutrality point, 
such that $\delta n \simeq \pm n$, one can use fitting to Eq.~(\ref{sigma-Csa})
in order to determine the concentrations $n_i^\pm$ of donors and acceptors 
separately, since the ratio 
$c=(\sigma_{\delta n = -n}-\sigma_{\delta n = n})
/(\sigma_{\delta n = -n}+\sigma_{\delta n = n})$.
Including the effect of the corrugations (ripples)\cite{corrugations}
simply shifts the symmetric part $n_i C_s \to n_i C_s + C_r$ 
in Eq.~(\ref{sigma-Csa}), 
where $\sigma_r \equiv (e^2/h)n/C_r$ is the conductivity due to the ripples only. 
This still allows one to determine the donor-acceptor imbalance $n_i^+ - n_i^-$.
To find $n_i^+ + n_i^-$ one needs to independently 
gauge the ripple parameter $C_r$ by varying the number of the charged impurities.
As expected, the Born approximation
$C_s\to C^{\rm Born} = \pi\alpha^2/2$ has no asymmetry
($C_a\to 0$), thus it entirely misses the effect.

Fig.~\ref{fig:Csa} shows the odd and even components (\ref{Csa})
of the exact cross-section, together with their polynomial fits
\bea 
\label{C-fit}
C_a &\approx& 1.68\,\alpha^3 +  7.69\,\alpha^5 -21.4\,\alpha^7 \,, \\
C_s &\approx& 1.54\,\alpha^2 +  2.33\,\alpha^4 \,, \quad 
C_a/C_s \approx 1.37\,\alpha  -1.06\,\alpha^3 \,.
\non
\eea
For $|Z|=1$ impurities in SiO$_2$
($\alpha\approx 0.37$) the asymmetry $C_a/C_s\approx 0.46$ should be 
fairly pronounced.\cite{asym-decrease}
Recent dc measurements of graphene monolayers on SiO$_2$ substrate
\cite{novoselov} show a noticeable asymmetry in the conductivity $\sigma(n)$. 
The proposed explanation is that the acceptors prevail over donors.
This provides an alternative to the suggestion
\cite{hwang-adam-dassarma} that the asymmetry is caused by the gate-induced
displacements of impurities. Further experiments with controlled numbers
of adsorbed donors and acceptors may distinguish between the two scenarios.

To conclude, a novel method for characterizing the electrostatic
environment of a graphene sample is proposed. The method is based on 
the attraction-repulsion asymmetry of the exact transport cross-section 
for scattering off the charged impurities, and could potentially allow 
to separate the scattering effects of the donors and of the acceptors from 
those of the microscopic ripples.

I thank L. Glazman, B. Shklovskii and A. Shytov 
for helpful discussions.
This work was supported by 
DOE Grant DE-FG02-06ER46310.



\begin{thebibliography}{99}
\bibitem{approach}
As $l_j$ is the impact parameter for the particle with the angular momentum 
$j= k_F l_j$, the asymptotic form of the scattering states\cite{novikov} 
(defining $\delta_j$) becomes valid for $r\gg l_j$.
\bibitem{stirling}
The sum in Eq.~(\ref{lambda-tr-C}) converges as 
$\alpha^2\sum_j (1/j^2)$ for $j\gg 1$, 
since $\delta_j \sim -\alpha_\epsilon \ln j$ according to the Stirling's formula.
\bibitem{asym-decrease}
This value may decrease somewhat when the additional 
screening by the free carriers is taken into account.
\end{thebibliography}
\end{document}